\documentstyle[12pt,aaspp4]{article} 
\newcommand{\etal}{{\it{et al.}}~}
\newcommand{\ie}{{\it{i.e.}}~}

\newcommand{\Ha}{H$\alpha$~}
\newcommand{\pccm}{pc cm$^{-6}$}
\newcommand{\arcs}{$^{\prime\prime}$}
\begin{document} 

\title{Diffuse Ionized Gas in Edge-on Spiral Galaxies: Extraplanar and
Outer Disk H$\alpha$ Emission\footnote{Observations made with the
Burrell Schmidt of the Warner and Swasey Observatory, Case Western
Reserve University.}}

\author{Charles G. Hoopes\altaffilmark{2}, Ren\'e
A. M. Walterbos\altaffilmark{2}}

\affil{New Mexico State University, Department of Astronomy, MSC 4500,
Box 30001\\ Las Cruces, New Mexico 88003 \\
choopes@nmsu.edu,rwalterb@nmsu.edu} 

\author{Richard J. Rand}

\affil{University of New Mexico, Department of Physics and
Astronomy, 800 Yale Boulevard NE\\ Albuquerque, New Mexico 87131-1156
\\rjr@astro.phys.unm.edu }

\altaffiltext{2}{Visiting Astronomer, Kitt Peak National Observatory,
National Optical Astronomy Observatories, which is operated by the
Association of Universities for Research in Astronomy, Inc. (AURA)
under cooperative agreement with the National Science Foundation.}

\begin{abstract}

We present H$\alpha$ images of five edge-on galaxies: NGC 891, NGC
4631, NGC 4244, NGC 3003, and UGC 9242. We also analyze [SII]
6717+6731\AA\space and [OIII] 5007\AA\space images of NGC 4631. For
several of these galaxies these images are the most sensitive to
date. We analyze the ionized gas content, with particular attention to
the diffuse ionized gas (DIG). The DIG layer in NGC 891 is traced out
to at least 5 kpc from the midplane, confirming an earlier
spectroscopic detection.  The DIG in four of these galaxies
contributes 40 to 50\% of the total \Ha luminosity, similar to face-on
galaxies, but in NGC 891 the DIG contributes 80 to 90\%. This is
likely due to the higher dust content in the disk of NGC 891 which
obscures the HII regions, but may also reflect the extraordinary
prominence of the DIG layer in that galaxy. Our very deep image of UGC
9242 shows very low surface brightness emission, as low as 0.3 \pccm,
reaching 4 to 5 kpc above the midplane. This galaxy also exhibits
filaments near the bright \Ha nucleus, an indication of a starburst
superwind. In NGC 4631 we see a very large shell of emission extending
3.5 kpc into the halo. The [SII]/\Ha and [OIII]/\Ha ratios in NGC 4631
are consistent with the ratios seen in other galaxies, and with
photoionization models. There is a region on the SE side of disk where
the [OIII]/(\Ha+[NII]) ratio reaches over 1.0 in the DIG, which
coincides with an HI supershell. We use our very deep images of NGC
3003 and UGC 9242 to search for ionized gas in the outer disks as a
test of the strength of the metagalactic ionizing radiation field. We
find no outer disk emission down to our 1$\sigma$ limit of 0.13
\pccm\space on scales of 1.5 kpc in NGC 3003. Based on this limit we
rule out a metagalactic ionizing radiation field stronger than
11$\times$10$^{-23}$ergs cm$^{-2}$s$^{-1}$Hz$^{-1}$sr$^{-1}$. There is
an indication of extended disk emission in UGC 9242 which would imply
a stronger radiation field, but various concerns, most importantly
flatfielding uncertainties due to foreground stars in the image, lead
us to question whether this feature is real.

\end{abstract} 


\section{Introduction}

Spiral galaxies contain a widespread layer of ionized Hydrogen, known
as diffuse ionized gas (DIG, also called WIM for warm ionized
medium). The properties of this component of the interstellar medium
(ISM) are important for many aspects of galactic research, such as the
influence of massive stars on the ISM, the porosity of the ISM, and
the disk-halo connection. The Reynolds layer, as DIG in the Milky Way
is known, has a filling factor of at least 0.2, and accounts for
nearly all of the mass of ionized gas, equal to about 30\% of the HI
mass. DIG is less dense than HII regions ($n_e$ $\sim$ 0.2 cm$^{-3}$,
compared to 10$^2$ to 10$^4$ cm$^{-3}$ in HII regions) but has a
similar temperature ( $\sim$ 8000K). An excellent review of the
properties of the Reynolds layer can be found in Reynolds
(\markcite{re90}1990).

The [SII] 6717+6731\AA\space to \Ha ratio is higher in the DIG
relative to HII regions, while [OIII] 5007\AA\space to \Ha is
lower. In the Reynolds layer and in M31 the [NII] 6548+6584\AA\space
to \Ha ratio is about the same as in HII regions
(\markcite{re89}Reynolds 1989; \markcite{gwb97}Greenawalt, Walterbos,
\& Braun 1997), while for NGC 891, the best studied edge-on galaxy, it
appears to be higher in the DIG (\markcite{ra97a}Rand 1997a). These
ratios can result from photoionization by a radiation field with a low
ionization parameter $U$, the ratio of the photon density to the gas
density (\markcite{m86}Mathis 1986; \markcite{dm94}Domg\"orgen \&
Mathis 1994). However, transporting ionizing photons from HII regions
to the DIG is still a problem, because of the long pathlengths that
ionizing photons have to travel. In order to remain ionized, the
Reynolds layer requires at least 15\% of the Lyman continuum photons
from OB stars in the Galaxy, equal to the amount of energy produced in
supernovae.  Studies of external galaxies suggest the requirement is
even higher, consistently 30 to 50\% (\markcite{wb94}Walterbos \&
Braun 1994;\markcite{hwg96} Hoopes, Walterbos, \& Greenawalt
1996;\markcite{fwgh96} Ferguson \etal 1996; \markcite{gwth98}
Greenawalt \etal 1998). Whether this energy is leaking out of
density-bounded HII regions, or can be provided by field OB stars is
still an open question. Recent spectroscopic observations have
challenged the photoionization models, including lower HeI
5876\AA\space than predicted (\markcite{rt95}Reynolds \& Tufte
1995,\markcite{ra97b} Rand 1997b) indicating a softer ionizing
spectrum than expected, higher [OIII] 5007\AA\space in some galaxies
than in others (\markcite{whl97}Wang, Heckman, \& Lehnert 1997;
\markcite{m97}Martin 1997) and rising [OIII]/\Ha ratio with height in
the halo of NGC 891 (\markcite{ra98}Rand 1998). Shock ionization by
supernovae is another source which likely plays a role at some level
and may explain the anomalous [OIII] ratios, but it cannot provide
enough energy to ionize the bulk of the DIG.

The Reynolds layer is vertically extended, with a scale height of 900
pc (\markcite{re90}Reynolds 1990). This distribution is quite
different from that of OB stars, which have a scale height closer
to 100 pc. \Ha imaging of external edge-on galaxies has revealed the
presence of DIG with varying properties. The first galaxy to be
studied was NGC 891, which has a very thick DIG layer that has been
traced as far as z=3.5 kpc in imaging (\markcite{rkh90}Rand, Kulkarni,
\& Hester 1990, hereafter RKH; \markcite{d90}Dettmar
1990; \markcite{pbs94}Pildis, Bregman, \& Schombert 1994), and z=5 kpc
with spectroscopy (\markcite{ra97b}Rand 1997b).  Several other
galaxies possess smaller DIG layers, and others show very little
extraplanar emission (\markcite{rkh92}Rand, Kulkarni, \& Hester 1992;
\markcite{ra96}Rand 1996; \markcite{w91}Walterbos 1991). Filamentary
structure is visible in the extraplanar emission of several galaxies,
indicating an active disk-halo connection and illustrating a probable
relation between DIG and star formation.

Here we present deep \Ha images of 5 edge-on galaxies. Some properties
of these galaxies are given in Table 1. We attempt to address several
questions with these images. First we would like to know the extent of
the DIG layers in galaxies down to levels fainter than previously
studied.  The DIG in NGC 891 may extend even farther into the halo at
very low surface brightness, and galaxies that do not show a bright
DIG layer may in fact have faint extraplanar emission. This is
important for understanding the structure of galaxies and their
gaseous halos.  Second, the correlation of DIG properties with other
parameters of galaxies, such as star-formation rate or Hubble type,
can be a clue toward the ionization source of the DIG, as well as the
mechanism for cycling gas into the halo. The strengths of other
emission lines can also provide information on the ionization source
through comparison with models (\ie \markcite{m86}Mathis 1986;
\markcite{dm94}Domg\"orgen \& Mathis 1994). For this we obtained [SII]
and [OIII] images NGC 4631.

Another goal of this project is to investigate the possibility of
detecting \Ha emission from the outer disks of galaxies. Such emission
is expected to occur when neutral Hydrogen in the outer disk and halo
is ionized by the metagalactic radiation field (\markcite{ss76}Silk \&
Sunyaev 1976).  The observation of a sharp cutoff in the HI disks of
several galaxies (\markcite{vg91}NGC 3198 by Van Gorkom 1991;
\markcite{css89}M33 by Corbelli, Schneider, \& Salpeter 1989) provides
indirect evidence that the outer disk gas may be ionized. Maloney
(\markcite{m93}1993), Dove \& Shull
(\markcite{ds94}1994) and Corbelli \& Salpeter (\markcite{cs93}1993)
have modeled the situation, and they predict the emission to be very
faint, from 0.2 \pccm\space in emission measure down to 0.05
\pccm\space (\markcite{m93}Maloney 1993). Fabry-Perot observations have 
pushed the observational limits close to the theoretical predictions
(Bland-Hawthorn, Freeman, \& Quinn 1997, Vogel \etal 1995). We
obtained very deep images of two edge-ons, NGC 3003 and UGC 9242, for
the purpose of searching for emission from the outer disk. Direct
detection of ionized gas would allow the determination of the strength
of the metagalactic radiation field at wavelengths below 912 \AA, and
would have important implications for cosmological models, for
prospects of measuring rotation curves of galaxies at large radii, and
for our knowledge of the structure of spiral galaxies.

The layout of this paper is as follows. In section 2 we detail the
observations and data reduction techniques. In section 3 we describe
the \Ha morphology of the galaxies in our sample.  In section 4 we
examine the DIG in these galaxies, including the vertical extent, and
contribution to the total \Ha luminosity. In section 5 we discuss the
[OIII] and [SII] images of NGC 4631. In section 6 we discuss the outer
disk emission in NGC 3003 and UGC 9242. Section 7 contains a
discussion of the implications of our results.

\section{The Data}

\subsection{Observations and Data Reduction}
 
The images discussed in this paper were obtained during several
different observing runs at KPNO. A log of the observations is given
in table 2. For all of the datasets we removed the bias level and bias
structure using standard methods in IRAF. Twilight flatfields were
combined into a super flatfield which was used to remove gain
variations. In some cases (noted below) the galaxies were observed
using the shift and stare technique, where the telescope is moved
between exposures to place the galaxy on different regions of the
chip. This allows the use of the image frames to produce a night sky
flatfield, if the galaxy is small enough in the field of view of the
telescope. We constructed a night sky flatfield by median combining
the galaxy images after editing out the galaxy and foreground
stars. The resulting image was heavily smoothed and then applied to
the original images after flattening with twilight flatfields.

NGC 891 was observed with the Burrell-Schmidt telescope at KPNO in
both \Ha and narrow-band continuum filters. Four pointings were
observed and combined into a mosaic. The large field of view of the
Schmidt (1$^{\circ}$.15) allowed us to make a night sky flatfield even
though NGC 891 is a relatively large galaxy (12$^{\prime}$). The final
flatfielding accuracy is about 1\% across the images. NGC 891 was also
observed with the 0.9 meter at KPNO, in \Ha and narrow-band
continuum. Two pointings (out of 4 planned) were observed. NGC 891 is
too big in the 23$^{\prime}$ field of view of the 0.9 meter to use the
images to make a night sky flatfield, but five blank sky images were
taken through the \Ha filter, with the telescope moved slightly after
each exposure. These were combined to make a night sky flatfield which
was then applied to the \Ha images.  The continuum image was already
flat to better than 1\% without the use of a night sky flatfield.

We obtained \Ha and narrow-band continuum images of NGC 3003 and UGC
9242 using the 0.9m telescope at KPNO. All of the images used were
taken during photometric conditions. The final mosaics consist of nine
pointings for each galaxy. A few of the images of UGC 9242 had to be
discarded due to large diffraction spikes from a bright star off of
the image. The 23$^{\prime}$ field of view is large enough compared to
these galaxies that the object images could be used to make a night
sky flatfield, which provided a flatfielding accuracy for NGC 3003 of
about 0.5\%, and probably much better than this in the region the
galaxy covers.  The night-sky flat alone did not produce satisfactory
results for the images of UGC 9242, but a second night sky flatfield
made from observations of blank sky worked well on the remaining
structure. The flatfielding uncertainty after this correction was
about 0.5\%, and as with NGC 3003 it is better over smaller scales.

NGC 4244 was observed at the KPNO 0.9 meter, during non-photometric
conditions. Three 30 minute exposures in the same position were
combined to produce the final image.  NGC 4631 was also observed at
the KPNO 0.9 meter in non-photometric conditions. A different detector
was used, which provided a 6.6$^{\prime}$ field of view. Three
pointings were required to cover the 14.3$^{\prime}$ length of NGC
4631, and these were combined into a mosaic during the reduction. The
small field of view and necessity of mosaicing make it difficult to
match background levels, hence flatfielding uncertainties may be
larger than for the other targets. Each pointing consists of
approximately 1.4 hours, with two pointings overlapping in the central
field (roughly 7 arcminutes). Images of the central field were also
obtained in [OIII] and [SII] filters.

The 0.9 meter observations of NGC 891, NGC 3003, and UGC 9242 were
calibrated using observations of spectrophotometric standard
stars. The Schmidt observation of NGC 891 and the 0.9 meter
observations of NGC 4244 and NGC 4631 were calibrated using the
R-magnitude of the galaxy. This was scaled by the response and
transmission of the continuum filter and used to calibrate the
continuum image. Then the stars in the line image were scaled to the
continuum, taking into account the differences in the response of the
filters. The [OIII] image of NGC 4631 was calibrated in a similar
manner using the V magnitude of the galaxy.  For convenience, all of
the \Ha images were converted to emission measure as if no [NII] was
transmitted by the filter. The wide filter used on NGC 4631 contains
the [NII] lines, but the narrow filters used for the other galaxies
transmit very little [NII]. The [NII]/\Ha ratio has been observed to
vary from 0.6 to 1.1 in the DIG of NGC 891 (\markcite{ra97a}Rand
1997a). At the highest value, [NII] transmitted by the narrow filters
contributes only 10$-$20\% of the observed \Ha flux.  For NGC 4631 we
correct for [NII] where it is relevant.

\subsection{Continuum Subtraction}

One of the most uncertain procedures in analyzing emission line
images is the removal of the underlying stellar continuum.
Subtracting too little continuum will leave a faint background that
can be interpreted as diffuse line emission, while subtracting too
much continuum can remove a layer of real diffuse emission. These
concerns can be minimized by observing the continuum close in
wavelength to the line, but varying and unknown \Ha absorption
produced by the stars in the galaxy makes a perfect subtraction
difficult to achieve. 

As a first determination of the continuum scale factor, we measured
the fluxes of foreground stars in both images, and computed the factor
needed to make them equal (for galaxies that were calibrated using the
R magnitude, the fluxes are made equal during calibration). This
relies on the foreground stars being of similar spectral type as the
stars in the galaxy, which is not necessarily the case. We then
visually inspected the images to be sure that there are no negative
regions, and that there is no obvious component of the stellar
continuum left in the image. For all of the galaxies the scale factor
derived from foreground stars was satisfactory, except the [OIII]
image of NGC 4631. For this image we adjusted the scale factor to
achieve the best subtraction. To quantify the dependence of our
analysis on the continuum subtraction, we vary the scale factor by
$\pm$ 3\%. At this level it is usually obvious that the continuum is
incorrectly subtracted. Our results throughout this paper will include
the variation in the continuum subtraction in the total uncertainty.

\subsection{Scattered Light}

Another potential problem with analyzing faint emission is the
possibility of scattered light. Telescope optics produce halos around
point sources, which could be mistaken for diffuse emission. In
addition, dust within the galaxy itself may scatter light from HII
regions, producing a halo of scattered light around HII
regions. Walterbos \& Braun (\markcite{wb94}1994) concluded that
scattered light from within the galaxy is not a major component, based
on the spectrum of the DIG. The [SII] to \Ha ratio is high in the DIG
compared to HII regions, which would not be expected if the DIG were
actually scattered light since the lines are very close in
wavelength. Ferrara \etal (\markcite{fbdg96}1996) modeled the
scattering of HII region light by dust in the halo of NGC 891, and
found it contributed only 10\% of the DIG emission at 600 pc off the
plane.

Methods of determining the contribution of scattered light from optics
have been explored by Walterbos \& Braun (\markcite{wb94}1994) and
Hoopes \etal (\markcite{hwg96}1996), and methods
for correcting for scattered light have been used by
RKH\markcite{rkh90}. Following these methods, we determined the
possible contribution of scattered light from bright foreground stars
in all of our images. We determined the radius in which 95\% of the
energy from the star is encircled, shown in table 2.  Scattered light
from HII regions will appear as halos around the regions to this
radius, with only 5\% of the light scattered further. We find that DIG
extends further than this in all 5 galaxies, at a much higher level
than 5\%.

Scattered light can also affect the determination of the scale height
of emission above the plane. To correct for this we deconvolved the
images with the stellar point spread function (PSF). The PSF was
measured using 10 stars in both the line and continuum images (except
NGC 4631, where only 5 stars were suitable). The images were then
deconvolved with the PSF using the Lucy-Richardson deconvolution
algorithm with the task LUCY in IRAF. This was done before calibration
and continuum subtraction.  Vertical profile fitting was done on the
deconvolved images (see section 4.1).

\section{Results for Individual Galaxies}

The \Ha images are shown in figures 1$-$5.  The \Ha luminosities of
the galaxies are listed in table 3. The correction for Galactic
extinction (\markcite{bh84}Burstein \& Heiles 1984) is negligible for
all of the galaxies except NGC891. The value listed in the table has
been corrected; the luminosity before correcting for extinction is 2.8
$\times$10$^{40}$ erg s$^{-1}$. No correction for internal extinction
was made. The uncertainty given in the table is due to varying the
continuum subtraction by $\pm$ 3\%. Our \Ha luminosity for NGC 891
agrees well with Rand \etal (\markcite{rkh92}1992). Our flux for NGC
4631 is about 6\% higher than that given in Rand \etal
(\markcite{rkh92}1992), but our image contains [NII] emission as well,
so our \Ha flux is actually lower. There are no published luminosities
of NGC 3003, NGC 4244, and UGC 9242 with which to compare. NGC 4631
and NGC 4244 were observed during non-photometric conditions.

In this section we discuss the general appearance of the \Ha images.
The \Ha morphology of several of the galaxies in our sample have been
discussed in great detail elsewhere, so we will keep our description
brief. These include NGC 891 (\markcite{rkh90}RKH;\markcite{d90}
Dettmar 1990;\markcite{pbs94} Pildis, Bregman, \& Schombert 1994) and
NGC 4631 (\markcite{rkh92}Rand \etal 1992;
\markcite{gdd96}Golla, Dettmar, \& Domg\"orgen 1996). NGC 4244 was
discussed briefly by Walterbos (\markcite{w91}1991) and Walterbos \&
Braun (\markcite{wb96}1996). NGC 3003 and UGC 9242 have not previously
been imaged in H$\alpha$.

\subsection{NGC 891}

In figure 1 we show both the Schmidt and the 0.9 meter continuum
subtracted \Ha images of NGC 891. Vertically extended emission is
clearly present.  The DIG layer can be traced to more than 3 kpc away
from the plane in these figures. The emission seems fairly uniform,
but does appear brighter and more extended near two large HII regions
on the east side of the disk. In the higher resolution 0.9 meter image
the extraplanar emission begins to show filamentary structure,
although it is not as pronounced as in NGC 4631 (see below). There is
a correlation with the brightest star forming regions, with the
brightest DIG near bright HII regions in the inner disk. Rand
(\markcite{ra97a}1997a) noted that the bright filaments almost always
connect to an HII region in the disk.

\subsection{NGC 3003}

The morphology of this galaxy (figure 2) suggests that it may be
disturbed, in that the spiral arms appear asymmetric. There is a small
dwarf galaxy (or possibly a background galaxy) at the lower right edge
of figure 2, but other than this there are no close companions visible
in our images, which cover an area of 300 $\times$ 300 kpc at the
distance of NGC 3003. However, just outside of our field of view is
the small spiral NGC 3021. These two galaxies are separated by about
200 kpc in projected distance, and about 50 km s$^{-1}$ in velocity
(Tully 1988).  The total \Ha luminosity of 1.18 $\times$ 10$^{41}$ erg
s$^{-1}$ indicates that active star formation is ongoing. This \Ha
luminosity is about an order of magnitude higher than for the
starbursts NGC 253 and M82. In fact, of the galaxies observed by Young
\etal (\markcite{y96}1996), one of the largest sample of galaxies
observed in \Ha in the literature, only six of the 120 spirals have
higher observed \Ha luminosities than NGC 3003 (not corrected for
extinction). The \Ha image shows a bright nucleus, and many bright HII
regions, including a very bright region in the outer western part of
the disk. A projected spiral arm protrudes below the galaxy in the
image, with a bright HII region at the end. The \Ha emitting disk is
about 35 kpc across. The galaxy is not quite edge-on, so it is
difficult to make any statements about the vertical extent of the DIG,
though there is pervasive diffuse emission in the disk.

\subsection{NGC 4244}

This is a nearby galaxy (3.1 Mpc), so we can resolve a great deal of
detail in the ionized disk. The HII regions distribution is extended
in the vertical direction, a result perhaps of a weaker disk potential
(\markcite{o96}Olling 1996). There are a few HII regions as far as 700
pc above the midplane. DIG is clearly visible in the disk, spread
between several bright HII regions.  It is immediately obvious,
however, that an extensive DIG layer such as in NGC 891 is absent in
this galaxy. Some short filaments exist in the central regions of the
galaxy, but overall the DIG layer is confined to the disk. The galaxy
appears quiescent in H$\alpha$, and the low \Ha luminosity and low FIR
surface brightness imply that there is relatively little star
formation occurring. Most of the HII regions are small and faint,
except for two bright complexes on either end of the disk.  The
absence of long bright filaments, such as those seen in NGC 891
and NGC 4631, imply that the galaxy also lacks any visible disk-halo
interaction.

\subsection{NGC 4631}

The \Ha+[NII] image of this galaxy is shown in figure 4. An overlay of
\Ha and x-ray emission for this galaxy, based on our data, can be
found in Wang \etal (\markcite{wwsnb95}1995). There appear to be
significant flatfielding uncertainties on the south side of the disk,
near the bottom edge of the image, so we restrict our analysis to the
north side. The image shows a disturbed disk, most likely due to an
interaction with its companion galaxies NGC 4627 and NGC 4656. The
disk appears to be actively forming stars, and the disk-halo interface
is also very active in this galaxy. Rand \etal (\markcite{rkh92}1992)
pointed out two bright vertical ``worms'' of emission east of the
nucleus on the north side of the disk. We note that these worms are
connected to longer, fainter filaments, and the western worm appears
to curve back around toward the disk. There is an even larger loop of
emission which surrounds the two worms (indicated by the arrows in
figure 4). It begins just east of the easternmost worm and extends 3.5
kpc into the halo. Unfortunately there is a seam in the image where
the loop might reconnect to the disk. The surface brightness at the
top of the loop is about 12 \pccm. The large loop is barely visible in
the \Ha image presented by Rand \etal (\markcite{rkh92}1992, their
figure 5). What appears to be another filament about 3 kpc east of the
giant loop in our image is actually another seam. There is significant
extraplanar emission in the form of discrete features such as the
loops and worms, and a smoother component can be traced up to about 2
kpc on the north side. Donahue, Aldering, \& Stocke
(\markcite{das95}1995) detected a faint halo extending 16 kpc from the
plane of NGC 4631, with a maximum brightness of 0.69 \pccm\space per
square arcsecond. We could not confirm the detection because of the
smaller field of view of our observations. This emission is probably
related to the high star formation and disturbed nature of this galaxy
resulting from the tidal interaction, as it is too bright to be caused
by the metagalactic ionizing radiation field.

\subsection{UGC 9242}

This galaxy appears to be close to exactly edge on (figure 5). The
continuum image shows a very thin and symmetric disk. The \Ha emitting
disk is about 35 kpc across. It has a bright nucleus in \Ha and
several bright star forming regions. There are two plumes visible on
the north side of the nucleus. They can be traced about 1.8 kpc from
the midplane, with typical emission measures ranging from 20 just
above the disk to about 5 at the highest point. These are probably
associated with star formation in the nucleus, but it is not
known whether these are two separate outflows, or whether they are the
brightened edges of a conical outflow such as that seen in NGC 253
(\markcite{ham90}Heckman, Armus, \& Miley 1990). If it is a conical
outflow, it is 1200 pc wide at the top. There is also some structure
on the other side of the nucleus, which also may be 2 spurs of
emission, shorter and weaker than the counterparts on the north
side. This suggests a double sided outflow, with a morphology similar
to the ``H'' shaped filaments seen in NGC 3079
(\markcite{hkrd90}Hester \etal 1990; \markcite{vcb95}Veilleux, Cecil,
\& Bland-Hawthorn 1995) and NGC 4013 (\markcite{ra96}Rand 1996). The
bright nucleus is about 1 kpc across, with an \Ha luminosity in a
12\arcs (about 1 kpc) diameter aperture of 4.2 $\times$ 10$^{39}$ erg
s$^{-1}$, comparable to the brightest HII regions in most galaxies
(\markcite{k88}Kennicutt 1988), and in fact equal to the total \Ha
luminosity of NGC 4244. Except for these filaments there is very
little obvious extraplanar emission, and there is no evidence for
disk-halo interaction beyond the nucleus, but like all of the galaxies
in our sample the disk is filled with DIG.

\section{The Diffuse Ionized Gas}

\subsection{Vertical Extent}

Walterbos \& Braun (\markcite{wb96}1996) compared the appearance of
the DIG layers in three of the galaxies in our sample, NGC 891, NGC
4244, and NGC 4631. Figure 6 is a more exact comparison, incorporating
UGC 9242. The images are shown on the same spatial scale, brightness
scale, and logarithmic stretch. The comparison shows the range of DIG
morphologies. The most prominent example is the smooth, bright,
extended layer in NGC 891. The patchy, filamentary layer in NGC 4631
may be related to recently enhanced star formation as a result of an
encounter. The most common appearance of the DIG may be more similar
to the weaker layers in NGC 4244 and UGC 9242. The inclination of NGC
3003 hinders our analysis of the extraplanar emission.

Following \markcite{rkh90}RKH and Rand (\markcite{ra96}1996), we have
attempted to characterize the DIG as an exponential layer. We fit the
vertical profile averaged over the central 10 kpc to increase the
signal to noise. In order to avoid including light from HII regions in
the fit, we excluded emission from $|z| \le$ 300 pc. NGC 4244 has a
thicker disk of HII regions, so we excluded emission from $|z| \le$
500 pc. We tried fits using a single exponential function as well as
fits using two exponentials. The parameters of the best fits for three
of the galaxies are given in table 4. The vertical profile of NGC 3003
suggests that it may also possess faint extraplanar emission, but the
lower inclination of this galaxy make the detection uncertain, as such
emission may arise from the outer part of the disk. We do not address it
further here, nor do we attempt the analysis for NGC 4631 due to its
disturbed nature and the poor flatfielding of the image. Note that the
scale heights given are for the surface brightness (emission
measure). The electron scale height is twice the emission measure
scale height, assuming the emission is only H$\alpha$. If much [NII]
is included in the filter, a rising [NII]/\Ha ratio in the halo such
as those observed in NGC 891 (\markcite{ra98}Rand 1998) and NGC 4631
(\markcite{gdd96}Golla, Dettmar, \& Domg\"orgen 1996) could make the
scale heights appear larger than they really are. The electron scale
height for the Reynolds layer is 900 pc (\markcite{re90}Reynolds
1990).

Previous imaging of NGC 891 (\markcite{rkh90}RKH) has traced the
extraplanar DIG to as far as 3.5 kpc from the plane of the
galaxy. Spectroscopy (\markcite{ra97b}Rand 1997b) has traced the
emission even further, to at least 5 kpc. Figure 7 shows a vertical
profile of the deconvolved Schmidt \Ha image of NGC 891, averaged over
the central 10 kpc. The emission can be traced as far as 5 kpc off the
plane in both directions. Thus we confirm the spectroscopic
detection. There is a suggestion that the emission continues even
further, to as far as 7 kpc, on the west side of the disk. However,
foreground stars in this part of the image make this
uncertain. Although the two fits look quite similar in the linear
scaling, the logarithmically scaled plot shows that two exponential
functions better fit the faint component of the profile. For
comparison, the stellar thin disk of NGC 891 has a scale height of 425
pc, and the stellar thick disk has a scale height of about 1.9 kpc
from surface photometry (\markcite{m99}Morrison 1999;
\markcite{mmhsb97}Morrison \etal 1997).

Even averaged over 10 kpc the extraplanar emission in NGC 4244 is very
weak (figure 8). Using two exponential functions, we find one
component with a very small scale height, about a factor of 2$-$3
lower than that of the Galaxy, and a fainter, broader component. The
faint component may be due to a flatfielding problem, such as low
level vignetting in the image. If we fit only one exponential the
scale heights are closer to that of the Galaxy (450 pc). The
two-exponential fit is statistically better, but both functions
describe the emission fairly well. NGC 4244 does not appear to possess
a thick stellar disk (\markcite{m99}Morrison 1999). 

Figure 9 shows that the extraplanar emission in UGC 9242 cannot be
described by a single exponential layer. Using two exponential
functions, the bright component of the vertical profile is well
described by a relatively low scale-height exponential, as expected
from the appearance of the \Ha image. However, there is a faint tail
of emission which is clearly visible out to 3$-$4 kpc. The scale
height of this component is similar to and perhaps even larger than
the scale height of the Reynolds layer. As noted, UGC 9242 has bright
filaments expending from the nucleus which may hamper the fit. In
figure 10 we show the vertical profile of the central 20 kpc with the
central 3 kpc excluded so as to remove the contribution of these
filaments. The parameters of the model fits are shown in table 4. A
single exponential still cannot fit the emission. The scale heights
for the bright component are similar to the 10 kpc fit, but the faint
components are even more extended. The high-z tail appears to extend
well past 5 kpc.

In figure 11 we show the profiles of NGC 891, UGC 9242, and NGC 4244
overplotted with the best fit model for each. Although NGC 891 has a
much brighter DIG layer than UGC 9242, extraplanar emission in both
galaxies actually reaches comparable distances above the disk. NGC
4244 clearly has no extraplanar emission comparable to the other two
galaxies. The logarithmically scaled plots clearly show that two
exponential components make up the extraplanar DIG in NGC 891 and UGC
9242, and possibly also in NGC 4244. 

Table 5 shows some parameters derived from the model fits. It is
necessary to know the diameter of the DIG cylinder to derive these
properties, so we estimate it from the \Ha images. The emission
measure in a column perpendicular to the disk is given for each of the
two components. We call the lower scale height component ``thick
disk'' DIG (so as not to be confused with the HII region thin disk),
and the higher scale height component ``halo'' DIG. The two components
contribute nearly equally to the total perpendicular emission measure
in NGC 891, while the halo components is weaker in the other two
galaxies. The Galaxy resembles UGC 9242 in terms of total
perpendicular emission measure and surface density, but keep in mind
that the Galactic DIG parameters are derived for the solar
neighborhood, while for the external galaxies these DIG parameters
apply closer to the center where it is brighter. Following
RKH\markcite{rkh90}, a constant filling factor of $\phi$=0.25 is
assumed when calculating the surface density, although it may rise in
the halo (\markcite{kh1988}Kulkarni \& Heiles 1988). The surface
densities include helium at solar abundance.

\subsection{Diffuse Fractions}

Another way to compare the DIG in different galaxies is through the
diffuse fraction, which is the contribution of the DIG to the total
\Ha luminosity. We us the same method used by Hoopes \etal 
(\markcite{hwg96}1996) for isolating diffuse emission from
HII region emission. This technique compensates for the varying
brightness of the DIG layer with radius in the galaxy and galaxy
inclination (the reasons a simple isophotal cut at a given surface
brightness fails). The method involves subtracting smoothed version of
the image from the original to remove the diffuse component, then
making a mask on the resulting image which removes pixels greater than
a certain value and replaces them with zero. The mask is applied to
the original image to remove HII regions emission. The technique was
tested by applying it the M31 images of Walterbos \& Braun
\markcite{wb92}(1992, \markcite{wb94}1994). We chose a scale of
900 pc for the median box used to smooth the image, and a cut level of
50 \pccm\space to make the mask, because these parameters resulted in
diffuse fractions similar to those that were found by manually
cataloging and removing the HII regions by hand in M31. Note that the
cut level of 50 \pccm\space in the image after subtracting the
smoothed version does not correspond to the same level in the original
image.

We applied this technique to the galaxies in our sample. The measured
diffuse fractions are given in table 3. Our \Ha flux NGC 4631 is low
compared to Rand \etal (\markcite{rkh92}1992), so to
test the effects of a possible calibration error we multiplied the
image by 1.3 and recalculated the diffuse fraction using the same
method, which gave 37 $-$ 41\%, very similar to the original
value. The diffuse fractions for NGC 3003, NGC 4244, NGC 4631, and UGC
9242 all fall within the 30 to 50\% range found for face on galaxies
(\markcite{hwg96}Hoopes \etal 1996;
\markcite{gwth98}Greenawalt \etal 1998). NGC 891, however, shows a
much higher fraction, 83$-$86\%. The diffuse fraction for NGC 891 was
measured on the 0.9 meter image. Applying the technique to the lower
resolution Schmidt image gives a fraction of 90\%. The difference
between the fractions measured on the Schmidt and 0.9 meter images of
NGC 891 is most likely due to unresolved HII regions in the lower
resolution Schmidt image being counted as DIG.

Comparison of the diffuse fractions found in edge-on galaxies to those
found in face-on systems is not straightforward. Since the HII region
layer is confined to the midplane where the dust density is the
highest, one might expect to measure a higher diffuse fraction in edge
on galaxies, as the HII regions would be affected more by the dust
disk than they would in face-on galaxies.  The DIG extends well out of
the dust disk, especially in NGC 891, so much of it is less
obscured. This may explain the high ratio in NGC 891. As the only Sb
galaxy in our sample, it may have a higher metal content than the
later type spirals, and the higher abundance of dust may obscure more
of the disk HII region emission. A comparison of the \Ha surface
brightness with the FIR surface brightness supports the idea that NGC
891 contains more dust than the other galaxies in our sample. The FIR
luminosity is an upper limit to the star formation rate, as an unknown
fraction arises from dust heated by older stars. The \Ha luminosity
gives a lower limit to the star formation rate, since it is affected
by extinction. Table 3 shows that the L$_{H\alpha}$/L$_{FIR}$ ratio
for NGC 891 is low relative to the other galaxies in the sample,
implying higher extinction. Interestingly, NGC 4244 has a high value
for this ratio, implying that it may be dust poor. 

\section{Line Ratios in NGC 4631}

Figure 12 shows a subsection of the \Ha image of NGC 4631 and the
[OIII]/\Ha and [SII]/\Ha ratio images. The figure shows a portion of
the disk east of the bulge of the galaxy. In the [OIII]/\Ha image the
cores of HII regions show a high ratio, while the DIG shows a lower
ratio. Not all HII regions have high [OIII]/\Ha ratios, however, and
in those that do only the central core has an elevated ratio, except
for one region described below. The [SII]/\Ha image shows the opposite
trend, with the DIG showing a higher [SII]/\Ha ratio than HII regions.

Applying the mask made on the \Ha image of NGC 4631 to the [OIII] and
[SII] images allows us to investigate the line strengths in the DIG
and in HII regions. The continuum in the bulge did not subtract well
from the [OIII] image, so we omitted that region and determined the
ratio in the disk east and west of the bulge.  The global [SII]/\Ha
and [OIII]/\Ha ratios are given in table 6.  The [SII]/(\Ha+[NII])
ratio is elevated in the DIG, while [OIII]/(\Ha+[NII]) is lower.  Note
that the \Ha filter also contains a contribution from the nearby [NII]
lines, which must be taken into account when comparing with other
measurements. In the disk [NII]/\Ha varies from 0.1 to 0.2 in HII
regions, and 0.3 to 0.5 in the DIG (\markcite{gdd96}Golla, Dettmar, \&
Domg\"orgen 1996). Thus when comparing to measurements made without
[NII], the observed [OIII]/\Ha and [SII]/\Ha ratios for NGC 4631 could
be 1.1 to 1.2 times higher in HII regions, and 1.3 to 1.5 times higher
in the DIG. Another important point is that the images are not
corrected for internal extinction. Greenawalt \etal
(\markcite{gwb97}1997) found that the extinction in HII regions is
higher than that in the DIG in M31. If this is also true in NGC 4631
then correcting for extinction would reduce the [OIII]/\Ha ratio in
HII regions more than in the DIG. The large uncertainties at high z
distance, due to low signal in the [OIII] and [SII] images, prevent us
from investigating the behavior of the line ratios away from the
plane.

There is an extended region on the SE side of the disk where the
[OIII]/\Ha ratio reaches over 1.0 (near the bottom of the images in
figure 12). This is comparable to the ratios seen in the cores of HII
regions, but the high [OIII] gas appears more extended than other HII
regions, and lies in a region where the \Ha emission is about 180
\pccm. By contrast, the cores of HII regions which show high [OIII]
have emission measures of about 1500 \pccm. This is the same location
in the galaxy where Rand \& van der Hulst (\markcite{rvdh93}1993)
found a large HI supershell. The shell has been modeled as a collision
of a high velocity cloud with the disk of NGC 4631
(\markcite{rs96}Rand \& Stone 1996). The region of high [OIII]
emission is about 650 pc in diameter, but borders on a large HII
region which also shows high [OIII]/\Ha, so the true extent is
difficult to measure. The ratios in a 370 pc wide vertical slice
through the region are shown in figure 13. The location of this slice
is marked in figure 12. The [SII]/\Ha ratio in this region is about
0.2, close to the value in HII regions. The gas here may be shock
ionized, which can produce a high [OIII]/\Ha ratio
(\markcite{sm79}Shull \& McKee 1979;
\markcite{ds95}Dopita \& Sutherland 1995). Shock-ionized gas can also 
have low [SII]/\Ha if the density is high enough to
collisionally de-excite S$^+$ (\markcite{ds95}Dopita \& Sutherland
(1995). This region was not included in the determination of the
global [OIII]/\Ha and [SII]/\Ha ratios.

\section{Limits on Emission from the Outer Disk}

One of the goals of this project was to detect or set limits on \Ha
emission from the outer disk. These observations were driven by the
recent discovery of sharp edges to the HI disks in nearby galaxies
such as NGC 3198 (\markcite{vg91}van Gorkom 1991). A possible
explanation for these edges is that the outer Hydrogen is ionized,
making it undetectable to 21 cm observations, as suggested by Silk \&
Sunyaev (\markcite{ss76}1976). There are no ionizing stars at these large
radii, and since the HI typically extends several kpc past the optical
or \Ha disk any ionizing radiation from within the galaxy would surely
be absorbed by the intervening HI. Therefore the ionizing source must
be extragalactic, and is thought to be the metagalactic ionizing
radiation field produced by quasars and AGN.

This idea has been modeled (\markcite{m93}Maloney 1993;\markcite{cs93}
Corbelli \& Salpeter 1993;\markcite{ds94} Dove \& Shull 1994) and the
observed cutoff can be reproduced. A crucial test, however, is to
directly detect the ionized outer disk. The models indicate that the
emission would be extremely faint, as low 0.05 \pccm, but possibly as
high as 0.2 \pccm, depending on the gas density, clumpiness, and the
strength of the ionizing radiation field. This is fainter than imaging
studies have reached in the past. Ionized gas was detected in the
outer disk of NGC 253 (\markcite{bfq97}Bland-Hawthorn \etal 1997) at a
level of 0.23 \pccm, using very sensitive Fabry-Perot
observations. The emission would imply a metagalactic radiation field
stronger than the current upper limit of 8$\times$10$^{-23}$ergs
cm$^{-2}$ s$^{-1}$ Hz$^{-1}$ sr$^{-1}$ (Vogel \etal
\markcite{vwrh95}1995), and the authors argue that the gas is
photoionized by disk OB stars which can see the warped outer disk.

With extreme care in flatfielding, it may be possible to detect this
level of emission through deep narrow-band imaging. Recently Donahue
\etal (\markcite{das95}1995) detected a very faint halo around NGC
4631. The surface brightness of this halo is as high as 0.69 \pccm,
much brighter than that expected from the metagalactic radiation
field, so the emission is most likely a result of the star formation
activity in the galaxy or related to the gravitational interaction
with its neighbors. We used similar imaging techniques in order to
optimize the detection of faint emission. Our target galaxies (NGC
3003 and UGC 9242) were chosen to be at high galactic latitude so that
they were not affected by many bright foreground stars or emission
from the Reynolds layer. The sensitivity of the NGC 3003 image is
about 2.0 \pccm\space per square arcsecond, and about 2.7 \pccm\space
per square arcsecond for UGC 9242, much brighter than the strongest
expected emission of 0.25 \pccm\space found by Maloney (1993). In
order to reach fainter levels, we spatially averaged over an ever
increasing area to lower the noise until the limit of the flatfielding
accuracy was reached and no further increase in S/N with smoothing was
apparent. The NGC 3003 image was binned into 25$\times$25 pixel
(17.3\arcs$\times$17.3\arcs) boxes, which increased the sensitivity by
a factor of 17.3 to 0.13 \pccm. The UGC 9242 could be averaged over
30$\times$30 pixel (20.7\arcs$\times$20.7\arcs) regions, increasing
the sensitivity to 0.13 \pccm. The rms intensity deviation between the
background levels in boxes in flat regions of the image was equal to
or less than this limit, even for boxes separated by large distances.
The spatial scale for these limits are 2 kpc for NGC 3003 and 2.6 kpc
for UGC 9242.

The binned images show no obvious outer disk emission. In figure 14 we
show the major axis profiles of the two galaxies from the binned
images. UGC 9242 has a distinct hole on the west side (positive major
axis distance in figure 14). This appears to be a flatfielding error
at the edge of one of the images which went into the final
mosaic. Aside from this, any outer disk emission is below the
limits of our flatfielding accuracy. We also smoothed using a median
filter, using the same size median box as the average box used
above. The NGC 3003 image shows no evidence for outer disk emission,
but the UGC 9242 image has some interesting features (see figure
15). Faint extraplanar emission is visible on both sides of the
disk. On the east side of the disk there is a bright star, which may
account for the emission in this region. On the west side, however,
the stars are fainter and less likely to contribute much scattered
light. On this side the emission ranges from about 2 \pccm\space at 2
kpc above the midplane to our limit of 0.13 \pccm\space at 8.5 pc
above the midplane. It appears that the emission is centered towards
the nucleus, which may imply a connection to the central starburst,
but might also just reflect the distribution of halo gas.

The \Ha disk also appears more extended in the median smoothed
image. The \Ha emission extends about 8 kpc past the southern edge of
the optical disk (the left side of figure 15). The optical edge is at
about -21 Mpc in figure 14. Several factors lead us to question the
validity of this feature. The emission ranges in brightness from 0.16
to 0.20 \pccm, just barely above our detection limit, making this a
1$-$2$\sigma$ detection at best, and is not confirmed in the binned
image. There is a background galaxy and some faint foreground stars
which may affect the flatfielding in this region. There are other
variations of similar magnitude near stars in the image. An example of
this can be seen just below the extended disk in figure 15, where
scattered light from a background galaxy and a group of several small
stars created a spot in the smoothed image. Although stars are removed
more cleanly during continuum subtraction in the deconvolved images
than in the original images, there are still residuals due to effects
such as a changing PSF across the field, leading to this further
source of uncertainty for faint emission.  If this emission were real,
it would contradict the limits set by the NGC 3003 image, as well as
other established upper limits on the metagalactic ionizing flux (see
section 7.2).

\section{Discussion}

\subsection{Halo Emission}

We have detected the DIG layer in NGC 891 out to at least 5 kpc from
the plane, and possibly as far as 7 kpc. NGC 891 has the brightest and
largest DIG layer known, and it is not surprising to detect emission
so far from the plane in a deep image. What is surprising is that UGC
9242, which shows much less extraplanar emission, still has a halo
extending as far as 3 to 4 kpc, though significantly fainter than the
NGC 891 halo. The image of UGC 9242 is of very high sensitivity, so it
is possible that similar faint halos might be seen around other
edge-ons even if they do not possess a bright DIG layer. We also
confirm the existence of extraplanar emission in NGC 4631, although we
could not have verified the halo claimed for NGC 4631 by Donahue \etal
(\markcite{das95}1995). For the other galaxies in our sample, even in
our very sensitive images of NGC 3003 and UGC 9242, we do not detect
such a halo.

Excluding NGC 3003, which has too low an inclination to properly study
the extraplanar emission, NGC 4631 and NGC 891 are the brightest in
H$\alpha$, and they also show the most extraplanar emission of the
sample. Although directly inferred star formation rates from the \Ha
luminosity can be very unreliable for edge-on galaxies due to
extinction, in a relative sense NGC 891 and NGC 4631 would appear to
be the most actively star forming galaxies of the sample (again
excluding NGC 3003). The extreme difference between these galaxies and
NGC 4244, with its low \Ha luminosity and weak DIG layer, point to a
link between active star formation and extraplanar emission, as
discussed previously by Rand (\markcite{ra96}1996).  This conclusion
is supported by the FIR luminosity (\markcite{ra96}Rand 1996), which
traces star formation more accurately than \Ha in edge-on galaxies,
although some fraction of the FIR emission may stem from dust heated
by general starlight, not OB stars. In table 3 we list the FIR
luminosity normalized by the square of the disk diameter (D$_{25}$),
following Rand (\markcite{ra96}1996). The galaxies with the highest
current star formation have the most prominent extraplanar
emission. This does not carry over to the diffuse fractions,
however. In fact the diffuse fractions seem relatively constant with
the exception of NGC 891, and this may be due to higher extinction in
that galaxy. NGC 891 is the only Sb in the sample and may have a
higher dust content than the rest of the sample (all Sc galaxies). The
constancy of the diffuse fraction further reinforces the connection
with star formation, since it essentially means that the DIG
luminosity scales with the HII region luminosity. It is also
interesting that there is not a more pronounced difference between
face-on and edge-on diffuse fractions. This might imply that we are
seeing most of the disk \Ha emission even in edge-on galaxies, with
the exception of NGC 891. Diffuse fractions for more edge-on spirals
are necessary to test this idea further.

The vertical profiles of NGC 891, UGC 9242, and possibly NGC 4244 are
best described by two distinct exponential components, raising the
possibility that more than one mechanism is responsible for creating
\Ha halos. NGC 891 possesses a thick stellar disk, while NGC 4244 does
not, which may suggest a connection between the mechanism which
creates a thick disk and that which is responsible for the \Ha
halo. However, a counter-example can be found in NGC 4565, which does
possess a thick stellar disk (\markcite{m99}Morrison 1999), but has
little extraplanar \Ha emission (\markcite{rkh92}Rand \etal 1992). An
x-ray observation of NGC 891 (\markcite{bp94}Bregman \& Pildis 1994;
\markcite{bh97}Bregman \& Houck 1997) revealed a halo of 10$^6$ K gas,
with a distribution similar to the \Ha emission. The idea that hot
supernovae-heated gas vented into the halo through chimneys is
responsible for this emission led the authors to calculate whether the
cooling of this hot gas could be the source of the \Ha emission. They
found that the cooled gas mass was an order of magnitude too low to
account for the mass of \Ha emitting gas. Rand (\markcite{ra97b}1997b)
suggested that cooling gas may explain the more-extended halo
component of the \Ha emission, while gas ionized by photons from OB
stars leaking out of the disk is responsible for the brighter,
less-extended component. The \Ha image of the DIG in UGC 9242 revealed
a very extended halo component, but the scale height of the disk
component is much lower than that of the similar component in NGC 891.
Perhaps cooling gas could be responsible for a faint halo such as in
UGC 9242, but some mechanism prevents the gas which makes up the disk
component from reaching as high as in NGC 891. There is indication of
an outflow into the halo from the nuclear filaments on both sides of
the disk, which might provide a source of hot halo gas. Table 6 shows
that the halo components in both UGC 9242 and NGC 4244 are much less
prominent relative to the disk components than is true for NGC 891,
where the two components are nearly equal. If cooling gas is
responsible for the halo component, NGC 891 must have a much more
active chimney mode than UGC 9242 and NGC 4244.

Recently several galaxies have been observed to have high [OIII]/\Ha
ratios in the DIG, including NGC 891 (\markcite{ra98}Rand 1998). The
implication is that a single ionization source cannot be responsible
for both the low [OIII] and high [OIII] emitting gas, so another
source of ionization is necessary. As Rand (\markcite{ra98}1998) and
Wang \& Heckman (\markcite{whl97}1997) point out, shock ionized gas
can have high [OIII]/H$\alpha$, so shocks may be an important
mechanism for ionizing this component of the DIG. In NGC 891 the high
[OIII] values are measured in the very high-z gas, suggesting that
shock ionization or some other mechanism is important in the upper
halo. In NGC 4631 we measure [OIII]/\Ha+[NII] ratios consistent with
the DIG in several external galaxies (\markcite{gwb97}Greenawalt \etal
1997; \markcite{whl97}Wang, Heckman, \& Lehnert 1997), but higher than
the Milky Way ratios. Photoionization models (\ie
\markcite{dm94}Domg\"orgen \& Mathis 1994) have been constructed which
reproduce the Galactic values, so they underpredict the [OIII] in NGC
4631. The [OIII]/\Ha ratio seen in the DIG of M31 (Greenawalt \etal
1997\markcite{gwb97}) is similar to those we find for NGC 4631. Those
authors found that a model adjusted so that the ionizing radiation is
less diluted on average could reproduce the observed ratios. A similar
situation may exist in NGC 4631, where the overall star formation is
enhanced in the disk. It then appears that photoionization is the only
ionization mechanism necessary for NGC 4631, in the disk at least. The
sensitivity at high z in the [OIII] and [SII] images is not good
enough to determine the ratio, so a trend such as that in NGC 891 may
well exist.

\subsection{Outer Disk Emission}

Our efforts to detect the outer disk \Ha emission allow us to set an
upper limit on the metagalactic ionizing radiation field. Maloney
(\markcite{m93}1993) constructed a model of the ionization of the
outer disk of NGC 3198 by the metagalactic radiation field. The model
predicts a range of expected \Ha surface brightness for different
values of the ionizing flux. An important result from the model was
that the predictions are insensitive to galaxy parameters, so we can
apply the results for NGC 3198 to the galaxies in our sample. The
models assume that the gas is smoothly distributed; if the gas is
clumpy the emission will be brighter.

Comparison of our upper limit on the \Ha emission of 0.13 \pccm\space
for the outer disk of NGC 3003 with the predicted emission measure in
\markcite{m93}Maloney (1993, see their figure 14) shows that the upper 
limit on the metagalactic ionizing flux J$_{\nu}$ is between 8 and 12
$\times$10$^{-23}$ergs cm$^{-2}$ s$^{-1}$ Hz$^{-1}$ sr$^{-1}$. The
calculations in
\markcite{m93}Maloney (1993) are based on the observed sharp cutoff in
the HI disk at large radii. The scenario is one in which the HI disk dips
below a critical column density where it becomes optically thin to the
metagalactic ionizing flux, going from a mostly neutral disk to a
mostly ionized disk.  Vogel
\etal (\markcite{vwrh95}1995) set a limit on extragalactic \Ha
emission using Fabry-Perot observations of an intergalactic Hydrogen
cloud. They placed a 2$\sigma$ limit on the metagalactic ionizing flux
to 8$\times$10$^{-23}$ergs cm$^{-2}$ s$^{-1}$ Hz$^{-1}$ sr$^{-1}$, in
agreement with the result presented here. Donahue \etal
(\markcite{das95}1995) found that
J$_{\nu}$$<$3.3$\times$10$^{-23}$ergs cm$^{-2}$ s$^{-1}$ Hz$^{-1}$
sr$^{-1}$, by imaging intergalactic Hydrogen clouds in H$\alpha$, more
stringent than our limit.

A simple check of these numbers can be made by calculating the number
of ionizing photons required to produce the observed \Ha emission. An
emission measure of 0.13 \pccm\space requires 8$\times$10$^{4}$
photons cm$^{-2}$ s$^{-1}$, assuming the HI is optically thick to
ionizing photons and that 45\% of ionizing photons result in \Ha
photons (Case B). The current limit is 6$\times$10$^{4}$ photons
cm$^{-2}$ s$^{-1}$ for two-sided incident ionizing flux (Vogel \etal
\markcite{vwrh95}1995). This would imply that our limit is a factor of
1.3 higher, so J$_{\nu}$ $<$ 11 $\times$10$^{-23}$ergs cm$^{-2}$
s$^{-1}$ Hz$^{-1}$ sr$^{-1}$. We have not quite reached the limits set
by Fabry-Perot observations nor that set by \Ha imaging of Hydrogen
clouds, so it is not surprising that we do not detect any outer disk
emission from NGC 3003. It is important to remember that the
conversion from \Ha emission measure to metagalactic ionizing flux
assumes that no [NII] is present. The filter used here may transmit
[NII] at the 10$-$20\% level. This is normally a small percentage of
\Ha, but it could be significant if the [NII]/\Ha ratio is very high
in the outer disk, as was found by Bland-Hawthorn \etal
(\markcite{bfq97}1997) for NGC 253.

In UGC 9242 there is an indication of outer disk \Ha emission at the
0.16 \pccm\space level. If this emission were a result of the
metagalactic ionizing radiation field, it would require that
J$_{\nu}$=17.0$\times$10$^{-23}$ergs cm$^{-2}$ s$^{-1}$ Hz$^{-1}$
sr$^{-1}$, higher than the Vogel \etal (\markcite{vwrh95}1995) upper
limits, and higher than the limit set by NGC 3003. The presence of
several foreground stars and a background galaxy at this location in
the image lead us to question whether this feature is real. It could
be either the result of flatfielding errors, or it could be real
emission but powered by some source other than the metagalactic
ionizing radiation field.

\acknowledgments

We thank the KPNO staff for their help during these observing runs. We
also thank B. Greenawalt for observing NGC 891, and M. F. Steakley for
reducing the NGC 4244 and NGC 4631 images. We thank the referee, James
Schombert, for useful comments which improved the presentation of the
results. This research was supported by the NSF through grant
AST-9617014, and by a Cottrell Scholar Award from Research
Corporation. C.G.H. was supported by a grant from the New Mexico Space
Grant Consortium.

\newpage
\begin{deluxetable}{lcccccccc}
\tablecolumns{9}
\tablewidth{0pc}
\tablenum{1}
\scriptsize
\tablecaption{Galaxy Parameters\tablenotemark{a}} 
\tablehead{
\colhead{Galaxy} & \colhead{Type} & \colhead{RA (2000)} & \colhead{dec
(2000)} &
\colhead{Distance} &
\colhead{inclination} & \colhead{D$_{25}$}\\ 
\colhead{} & \colhead{} & \colhead{} & \colhead{} &
\colhead{(Mpc)} &
\colhead{($^\circ$)} & \colhead{(arcmin)}} 
\startdata
NGC 891  & Sb & 2h19.3m & 42$^\circ$.07 & 9.5 & 89\tablenotemark{b} & 12.2  
\nl
NGC 3003 & SBbc & 9h45.6m & 2$^\circ$.51 & 24.4 & 90\tablenotemark{c} & 6.0 
\nl
NGC 4244 & Scd  & 12h15.0m & 38$^\circ$.05 & 3.1  & 90 & 15.8  \nl
NGC 4631 & Scd  & 12h39.8m & 32$^\circ$.49 & 6.9  & 85 & 14.7  \nl
UGC 9242 & Scd  & 14h23.3m & 39$^\circ$.45 & 26.3 & 90 & 5.0   \nl
\enddata
\tablenotetext{a}{From \markcite{t88}Tully 1988, except where otherwise noted.}
\tablenotetext{b}{From \markcite{r91}Rupen 1991.}
\tablenotetext{c}{The inclination of NGC 3003 appears to be less than
that listed in \markcite{t88}Tully 1988.}
\end{deluxetable} 

\begin{deluxetable}{lcccccc}
\tablecolumns{7}
\tablewidth{0pc}
\tablenum{2}
\tablecaption{Observations} 
\scriptsize
\tablehead{
\colhead{Galaxy} & \colhead{Filter\tablenotemark{a}} &
\colhead{Telescope} & \colhead{Date} & 
\colhead{Integration Time} & \colhead{RMS Noise
\tablenotemark{b}} & \colhead{PSF Radius \tablenotemark{c}} \\
\colhead{} & \colhead{} & \colhead{} & \colhead{} & \colhead{(Hours)} &
\colhead{(EM arcsec$^{-2}$)} & \colhead{(arcsec)}  
} 
\startdata
NGC 891  & \Ha6580/28 &  Schmidt & 1995 Nov & 3.3 & 2.3 & 8.1 \nl
 & \Ha6580/28  &  0.9m & 1996 Jan & 1.3 & 5 & 4.8 \nl
NGC 3003 & \Ha6590/28 &  0.9m & 1997 Mar & 10.8 & 1.6 & 4.1\nl
NGC 4244 & \Ha 6569/29  &  0.9m & 1992 Mar & 1.5 & 4.5 & 3.7 \nl
NGC 4631 & \Ha 6568/74 &  0.9m & 1991 Feb & 2.8 & 4 & 3.9 \nl
 & [SII] 6744/84  &  0.9m & 1991 Feb & 1.7 & 6 & 3.9 \nl
 & [OIII] 5024/55  &  0.9m & 1991 Feb & 1.8 & 11 & 4.7 \nl
UGC 9242 & \Ha6590/28 &  0.9m & 1997 Mar & 10.2 & 2.0 & 4.1 \nl
\enddata
\tablenotetext{a}{The name of the line is given, as well as the
central wavelength and FWHM in \AA.} 
\tablenotetext{b}{Refers to the continuum subtracted image.}
\tablenotetext{c}{The radius which encircles 95\% of the energy from a
point source (foreground star).}
\end{deluxetable} 

\begin{deluxetable}{lcccc}
\tablecolumns{5}
\tablewidth{0pc}
\tablenum{3}
\scriptsize
\tablecaption{\Ha Properties} 
\tablehead{
\colhead{} & \colhead{L$_{H\alpha}$\tablenotemark{a}} &
\colhead{Diffuse Fraction\tablenotemark{b}} &
\colhead{L$_{H\alpha}$/D$_{25}^2$} &
\colhead{L$_{FIR}$/D$_{25}^2$\tablenotemark{c}} \\
\colhead{Galaxy} & 
\colhead{(10$^{40}$ erg s$^{-1}$)} &
\colhead{} &
\colhead{(10$^{37}$ ergs s$^{-1}$ kpc$^{-2}$)} &
\colhead{(10$^{40}$ ergs s$^{-1}$ kpc$^{-2}$)}}
\startdata
NGC 891  & 3.3  $\pm$ 0.3 & 83-86\%  & 2.9  & 4.3      \nl
NGC 3003 & 11.8   $\pm$ 0.6  & 49-51\%  & 6.1  & 8.0    \nl
NGC 4244 & 0.42 $\pm$ 0.02  & 44-49\%  & 2.1  & 0.19 \nl
NGC 4631\tablenotemark{d} & 14.3 $\pm$ 0.2 & 38-42\%  & 16.4 & 3.5 \nl
UGC 9242\tablenotemark{e} & 2.8  $\pm$ 0.2  & 46-50\%  & 1.9  & \nodata  \nl
\enddata
\tablenotetext{a}{The uncertainty is found by varying the continuum 
subtraction by $\pm$3\%.}
\tablenotetext{b}{The range is found by varying the continuum
subtraction. The fractions are not corrected for internal extinction,
which may account for the large fraction in NGC 891.}
\tablenotetext{c}{IRAS FIR fluxes are from \markcite{ri88}Rice \etal 1988.}
\tablenotetext{d}{The \Ha filter used for NGC 4631 also contains [NII].}
\tablenotetext{e}{UGC 9242 was not detected by IRAS.}
\end{deluxetable} 

\begin{deluxetable}{lccccc}
\tablecolumns{6}
\tablewidth{0pc}
\tablenum{4}
\scriptsize
\tablecaption{Exponential Model Fits to the Vertical EM 
Profiles\tablenotemark{a}} 
\tablehead{
\colhead{}    &  \multicolumn{2}{c}{First Exponential} &
\multicolumn{2}{c}{Second Exponential} & \colhead{} \\
\colhead{z direction} & \colhead{EM$_{midplane}$\tablenotemark{b}} & 
\colhead{Scale Height} & \colhead{EM$_{midplane}$\tablenotemark{b}} &
\colhead{Scale Height} & \colhead{Reduced $\chi^2$}}
\startdata
\cutinhead{NGC 891, Central 10 kpc}
East  & 64  & 430 & 45 & 870  & 0.91 \nl
West  & 105 & 520 & 25 & 1350 & 0.93 \nl
East  & 87  & 710 & \nodata & \nodata & 1.02 \nl
West  & 94  & 850 & \nodata & \nodata & 1.59 \nl
\cutinhead{NGC 4244, Central 10 kpc}
North  & 180 & 130 & 7  & 450  & 0.92 \nl
South  & 40  & 200 & 2  & 860  & 0.77 \nl
North  & 25  & 300 & \nodata & \nodata & 1.11 \nl
South  & 10  & 450 & \nodata & \nodata & 1.14 \nl
\cutinhead{UGC 9242, Central 10 kpc}
North  & 230 & 180 & 10 & 1150 & 1.71 \nl
South  & 100 & 240 & 5  & 1700 & 1.20 \nl
North & 87  & 380 & \nodata & \nodata & 5.50 \nl
South & 44  & 500 & \nodata & \nodata & 5.08 \nl
\cutinhead{UGC 9242, Central 20 kpc, excluding nucleus}
North  & 300 & 180 & 3  & 1900 & 1.01 \nl
South  & 130 & 270 & 4  & 2400 & 1.07 \nl
North  & 160 & 250 & \nodata & \nodata & 5.55 \nl
South  & 75  & 410 & \nodata & \nodata & 7.11 \nl
\enddata
\tablenotetext{a}{For each galaxy, the first two rows give the results
for the model using two exponential functions, and the third and
fourth rows give the results for a single exponential function.}
\tablenotetext{b}{This is the emission measure of the line of sight
through the center of the disk (z=0), extrapolated from the vertical 
profile.}
\end{deluxetable} 

\begin{deluxetable}{lccccc}
\tablecolumns{6}
\tablewidth{0pc}
\tablenum{5}
\scriptsize
\tablecaption{Average DIG Parameters for Central 10 kpc} 
\tablehead{
\colhead{} & \colhead{DIG Diameter} & \colhead{Thick Disk
EM$_{\perp}$\tablenotemark{a}} & 
\colhead{Halo EM$_{\perp}$\tablenotemark{a}} 
& \colhead{Total EM$_{\perp}$\tablenotemark{a}} 
& \colhead{Surface Density} \\ 
\colhead{Galaxy} & \colhead{(kpc)} & \colhead{(pc cm$^{-6}$)} &
\colhead{(pc cm$^{-6}$)} &\colhead{(pc cm$^{-6}$)} & \colhead{(M$_{\odot}$ $
pc^{-2}$)}} 
\startdata
NGC 891   & 15 & 5.5 & 4.9 & 10.4 & 5.1($\phi$/0.25)$^{1/2}$ \nl
NGC 4244  & 11 & 2.9 & 0.5 & 3.4  & 1.5($\phi$/0.25)$^{1/2}$ \nl
UGC 9242  & 20 & 3.3 & 1.0 & 4.3  & 2.6($\phi$/0.25)$^{1/2}$ \nl
Galaxy\tablenotemark{b}    & \nodata & \nodata & \nodata & 4.5 & 2.6 \nl
\enddata
\tablenotetext{a}{Emission measure perpendicular to the disk, 
calculated by integrating the two-component exponential functions 
in table 4.} 
\tablenotetext{b}{From \markcite{re89}Reynolds 1989.}
\end{deluxetable} 

\begin{deluxetable}{lcc}
\tablecolumns{3}
\tablewidth{0pc}
\tablenum{6}
\scriptsize
\tablecaption{NGC 4631 Line Ratios in the Disk\tablenotemark{a}} 
\tablehead{
\colhead{Line} & \colhead{HII Regions} & \colhead{DIG} }  
\startdata
[SII]/(\Ha+[NII])  & 0.26 & 0.41 \nl
[OIII]/(\Ha+[NII])East & 0.42 & 0.26 \nl
[OIII]/(\Ha+[NII])West & 0.53 & 0.14 \nl
\enddata
\tablenotetext{a}{Not corrected for extinction.}
\end{deluxetable} 
\newpage

\clearpage
\begin{figure}
\figurenum{1}
\epsscale{0.7}

\caption{Left: The continuum-subtracted \Ha image of NGC
891, taken with the Burrell Schmidt. Right: A section of the KPNO
0.9 meter \Ha image of the same galaxy, showing the central 16 kpc of
the galaxy. North is up and east is to the left in this and all other
images, unless otherwise indicated. The bar in the lower left corner
is 1 kpc at the distance of NGC 891.}
\end{figure} 

\begin{figure}
\figurenum{2}
\epsscale{0.7}
\caption{Top: The red continuum image of NGC 3003. Bottom:
The continuum subtracted \Ha image of the same galaxy.  The bar in the
lower left corner is 2 kpc at the distance of NGC 3003}
\end{figure} 

\begin{figure}
\figurenum{3}
\epsscale{0.7}
\caption{The continuum subtracted \Ha image of NGC
4244. The bar in the lower left corner is 1 kpc at the distance of NGC
4244.}
\end{figure} 

\begin{figure}
\figurenum{4}
\epsscale{0.5}
\caption{The continuum subtracted H$\alpha$+[NII] image of
NGC 4631. The bar in the lower left corner is 1 kpc at the distance of
NGC 4631. The arrows indicate the large loop extending 3.5 kpc into
the halo.}
\end{figure} 

\begin{figure}
\figurenum{5}
\epsscale{0.7}
\caption{Top: The red continuum image of UGC 9242. Bottom:
The continuum subtracted \Ha image of the same galaxy. The bar in the
lower left corner is 2 kpc at the distance of UGC 9242.}
\end{figure} 

\begin{figure}
\figurenum{6}
\epsscale{0.7}
\caption{A comparison of the DIG layers in the galaxies in
our sample. The images have been rotated so the disk is horizontal;
see figures 1$-$5 for the correct orientation. The images all have the
same spatial scale, shown by the 1 kpc bar in the bottom panel. They
are all displayed with the same logarithmic stretch, from -2 to 1000
\pccm.}
\end{figure} 

\newpage
\begin{figure}
\figurenum{7}
\epsscale{1.0}
\plotone{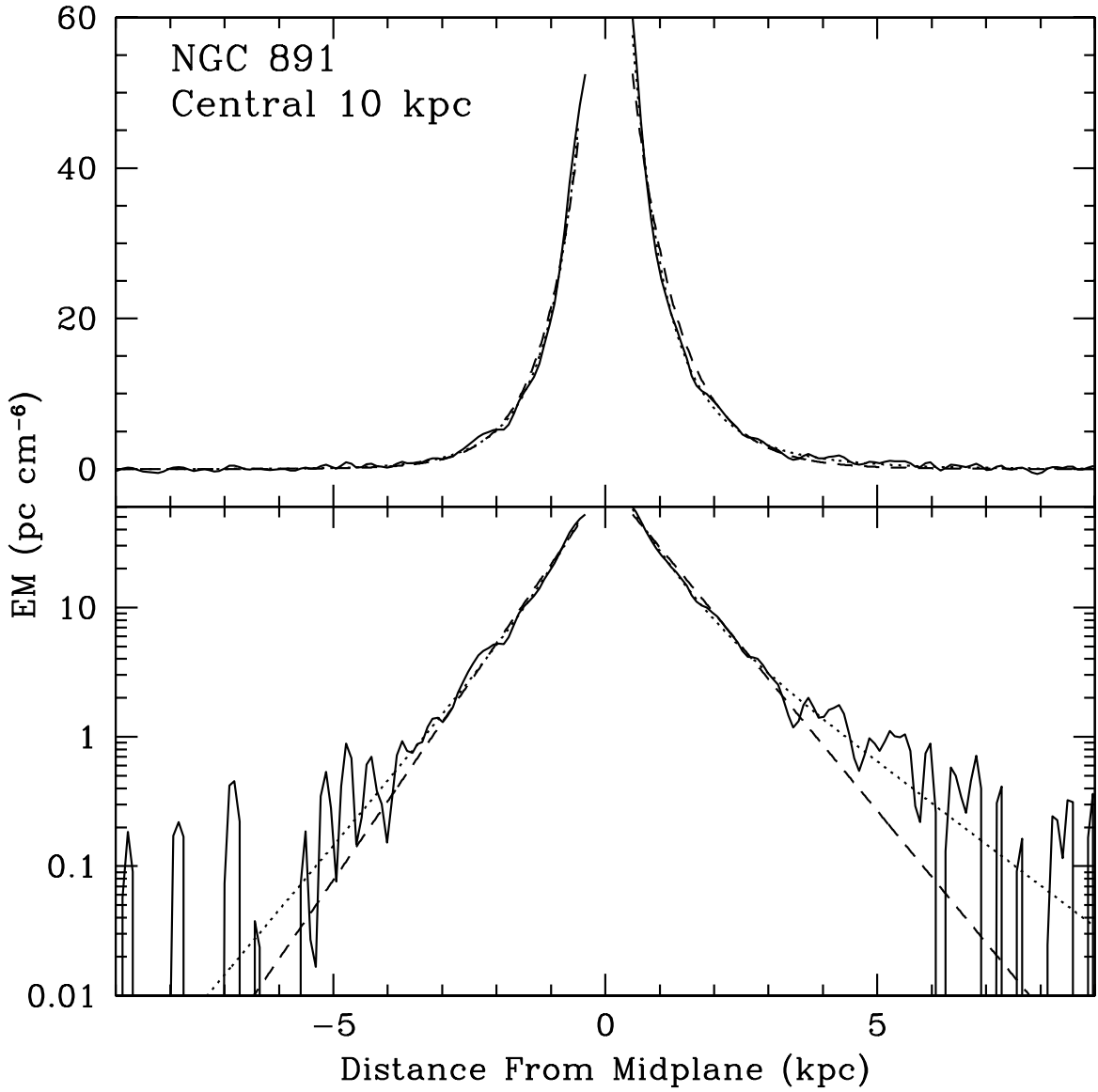}
\caption{Vertical profile of the central 10 kpc of NGC
891. The top panel is linearly scaled, and the bottom panel is
logarithmically scaled. Both the two-exponential fit (dotted line)
and the single exponential fit (dashed line) from table 4 are shown.}
\end{figure}

\newpage
\begin{figure}
\figurenum{8}
\epsscale{1.0}
\plotone{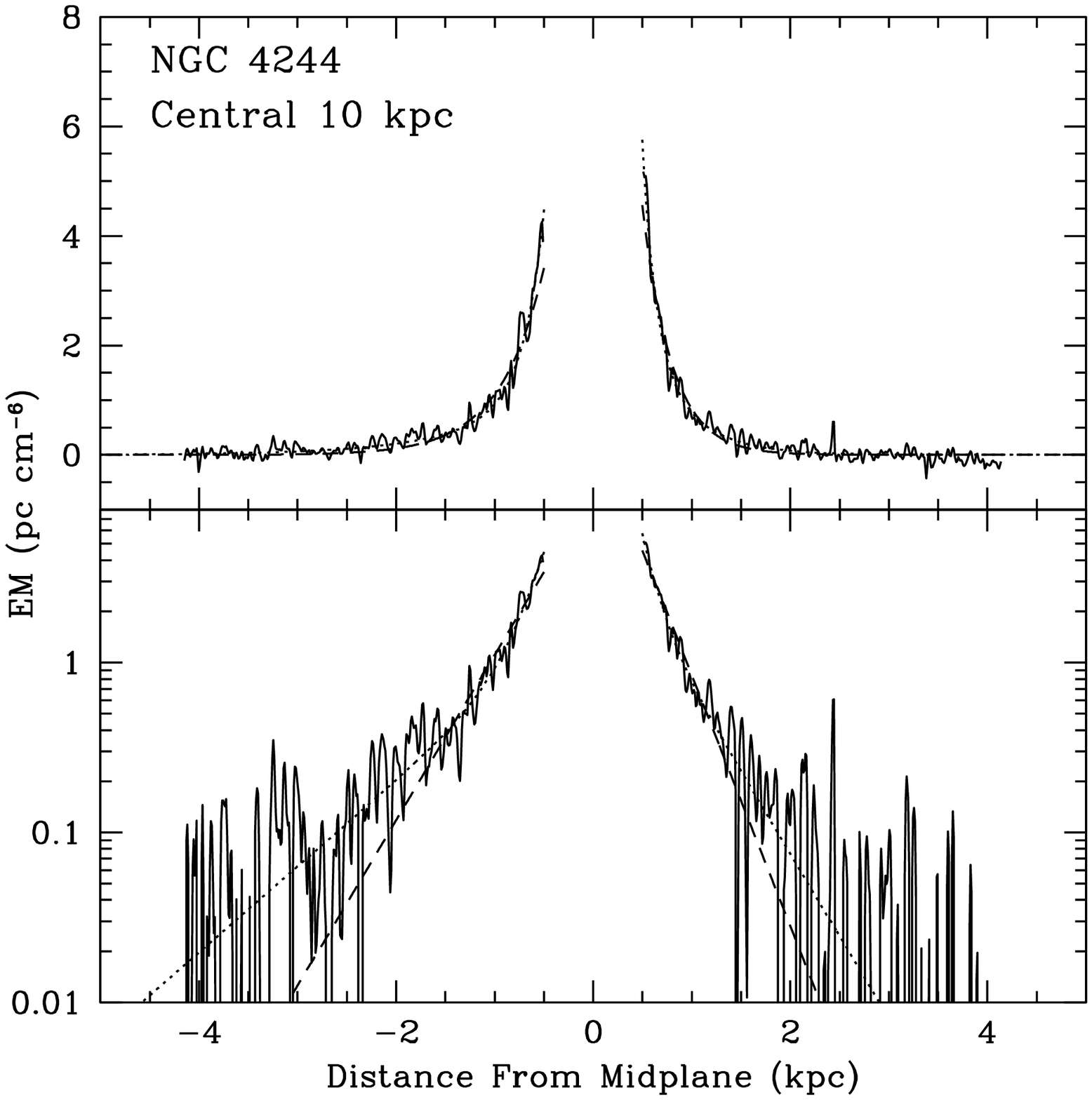}
\caption{Vertical profile of the central 10 kpc of NGC
4244. Both the two-exponential fit (dotted line) and
the single exponential fit (dashed line) from table 4 are shown.}
\end{figure}

\newpage
\begin{figure}
\figurenum{9}
\epsscale{1.0}
\plotone{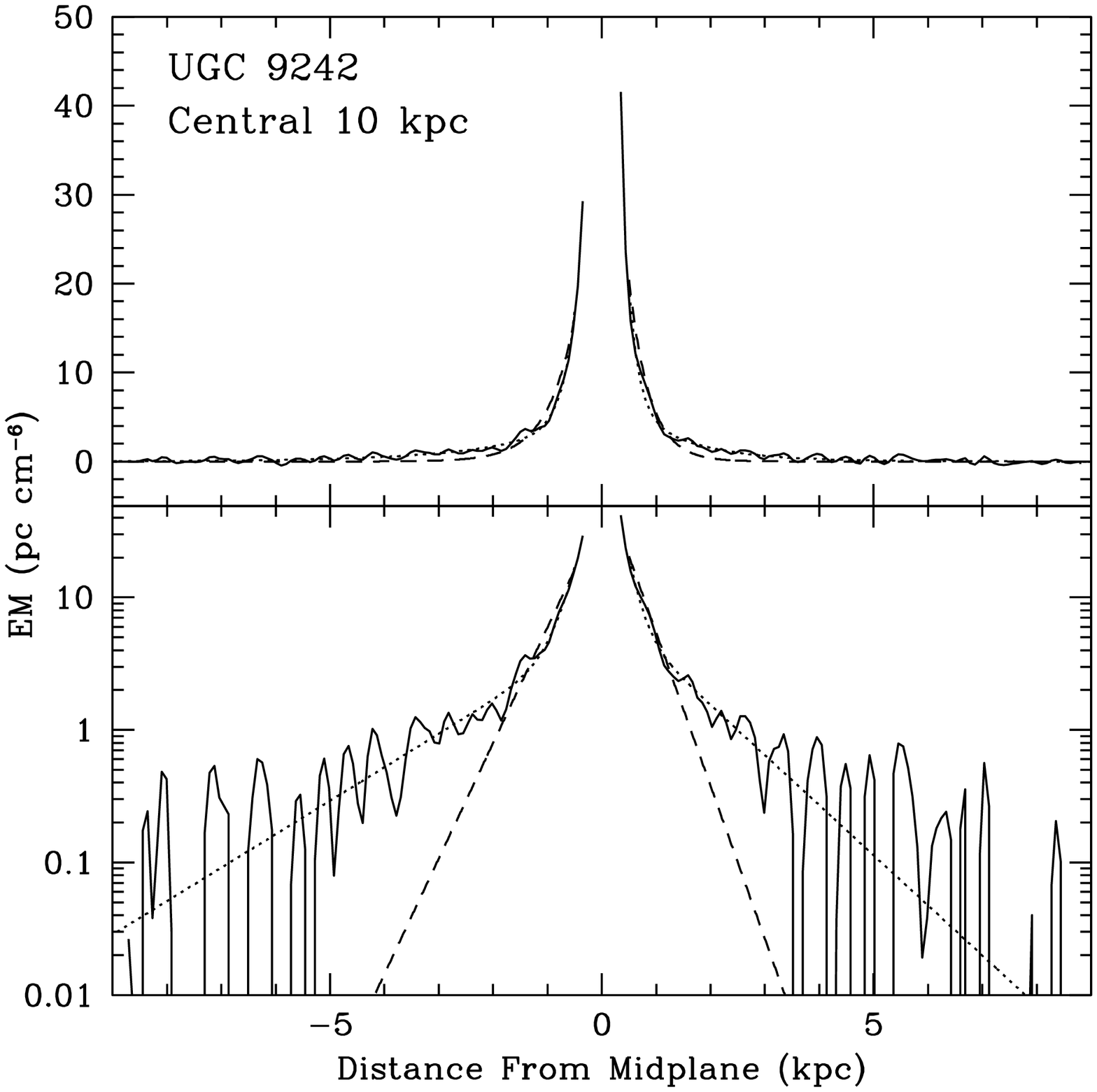}
\caption{Vertical profile of the central 10 kpc of UGC
9242. Both the two-exponential fit (dotted line) and
the single exponential fit (dashed line) from table 4 are shown.}
\end{figure}

\newpage
\begin{figure}
\figurenum{10}
\epsscale{1.0}
\plotone{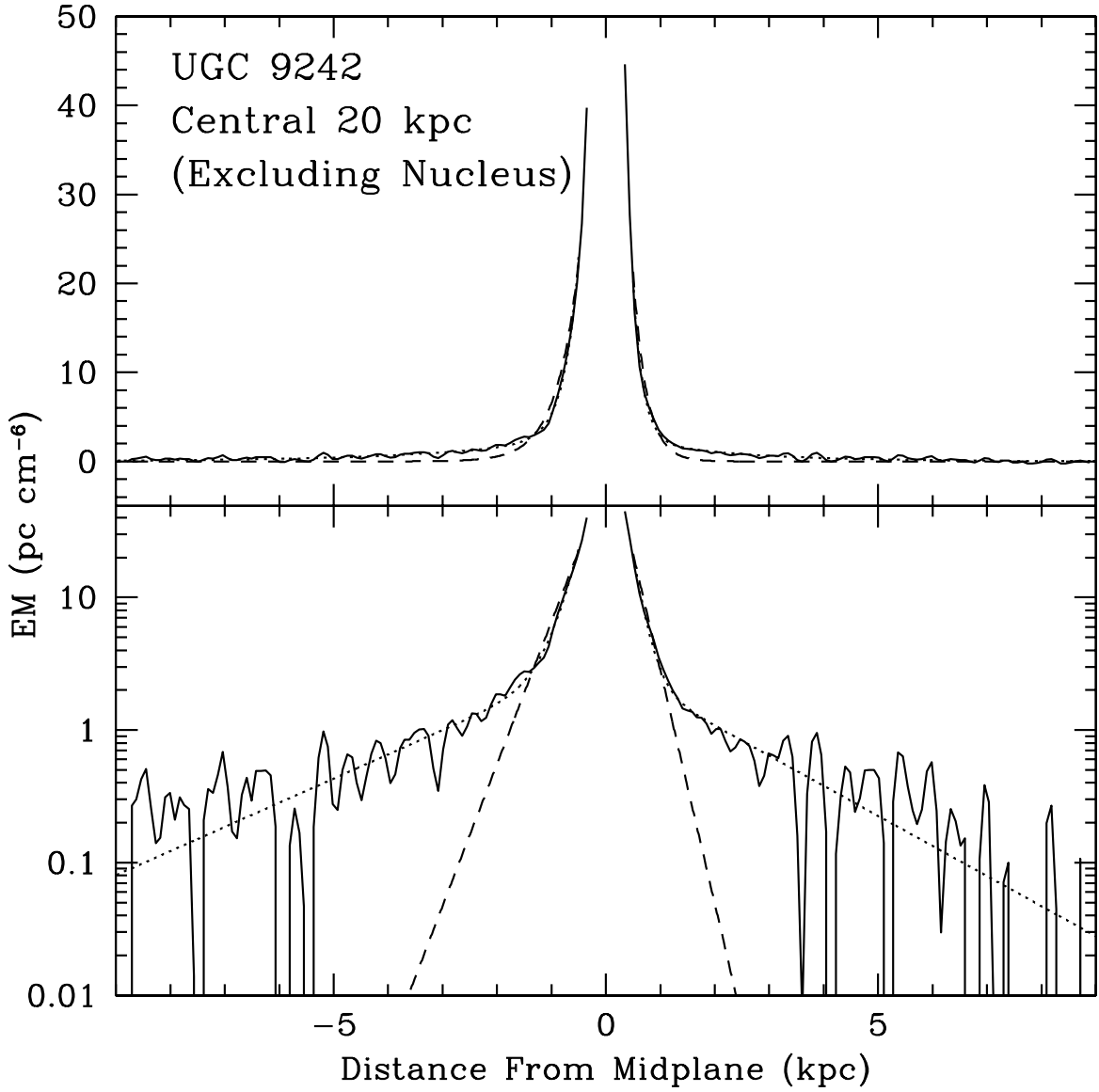}
\caption{Vertical profile of the central 20 kpc of UGC
9242, with the central 3 kpc excluded. Both the two-exponential fit
(dotted line) and the single exponential fit (dashed line) from table
5 are shown.}
\end{figure}

\newpage
\begin{figure}
\figurenum{11}
\epsscale{1.0}
\plotone{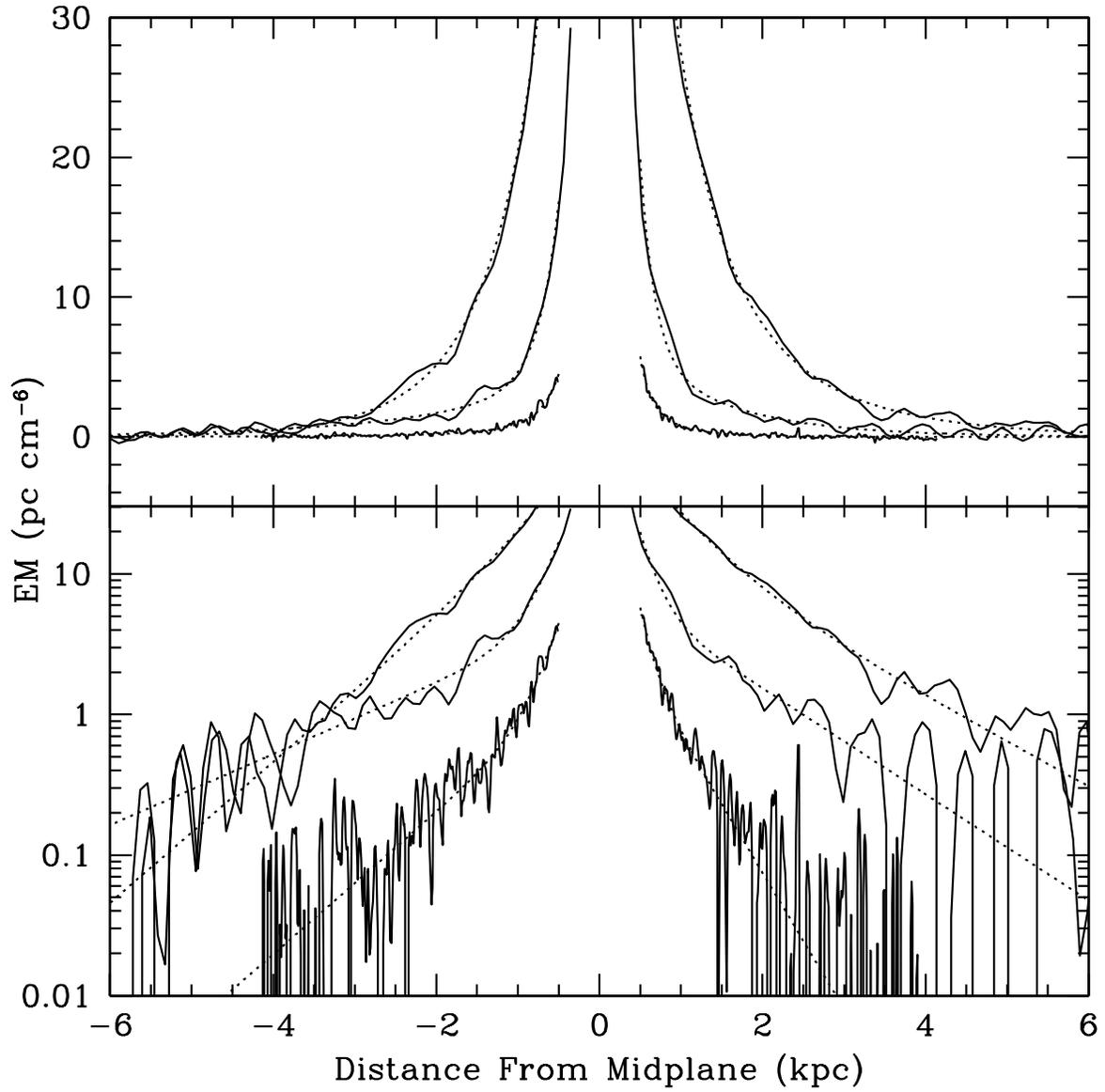}
\caption{Vertical profile of the central 10 kpc of all
three galaxies, shown in linear and logarithmic scale. The
two-exponential fits from table 4 (dotted line) are shown.}
\end{figure}

\newpage
\begin{figure}
\figurenum{12}
\epsscale{0.9}
\caption{A subsection of NGC 4631 showing the disk east of 
the bulge. The top panel is \Ha, the middle is [OIII]/\Ha, and the
lower panel is [SII]/\Ha. Note that the \Ha image also contains [NII]
emission. The [OIII]/\Ha image is scaled linearly from 0 to 1.2, and
the [SII]/\Ha image is scaled linearly from 0 to 0.6. White represents
high values in all three panels. The horizontal bar represents 1 kpc,
and the vertical lines show the region of these images plotted in
figure 13.}
\end{figure} 

\newpage
\begin{figure}
\figurenum{13}
\epsscale{1.0}
\plotone{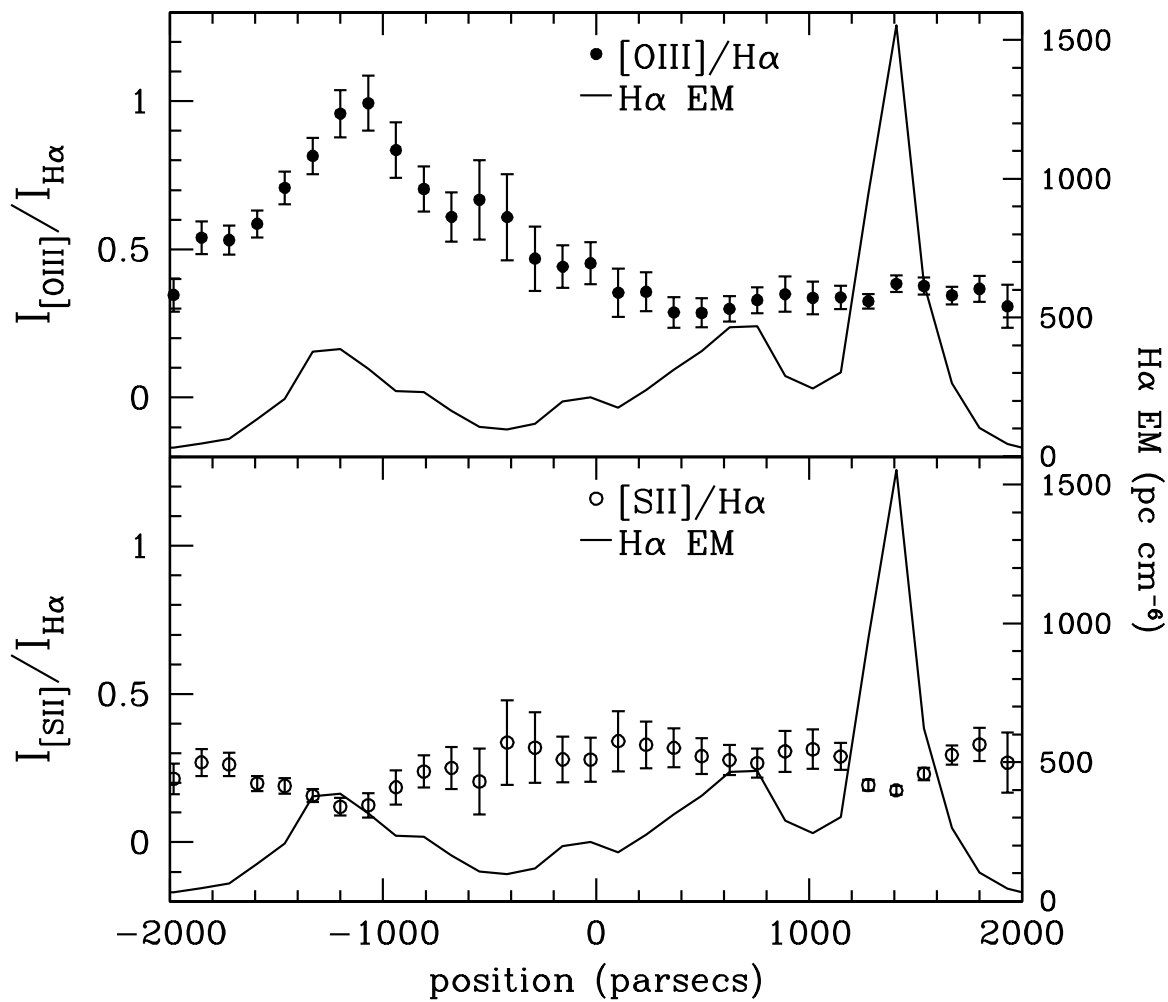}
\caption{The \Ha surface brightness (solid line), [OIII]/\Ha 
ratios (filled circles in the top panel) and [SII]/\Ha ratio (open
circles in the bottom panel) in a 370 pc wide vertical slice through
NGC 4631 (see figure 12). The slice passes through the high [OIII]
emitting gas which may be related to an HI supershell (see text). The
region is at about $-$700 to $-1500$ parsecs in this plot. Note that
the \Ha image contains [NII] emission.}
\end{figure}

\newpage
\begin{figure}
\figurenum{14}
\epsscale{1.0}
\plotone{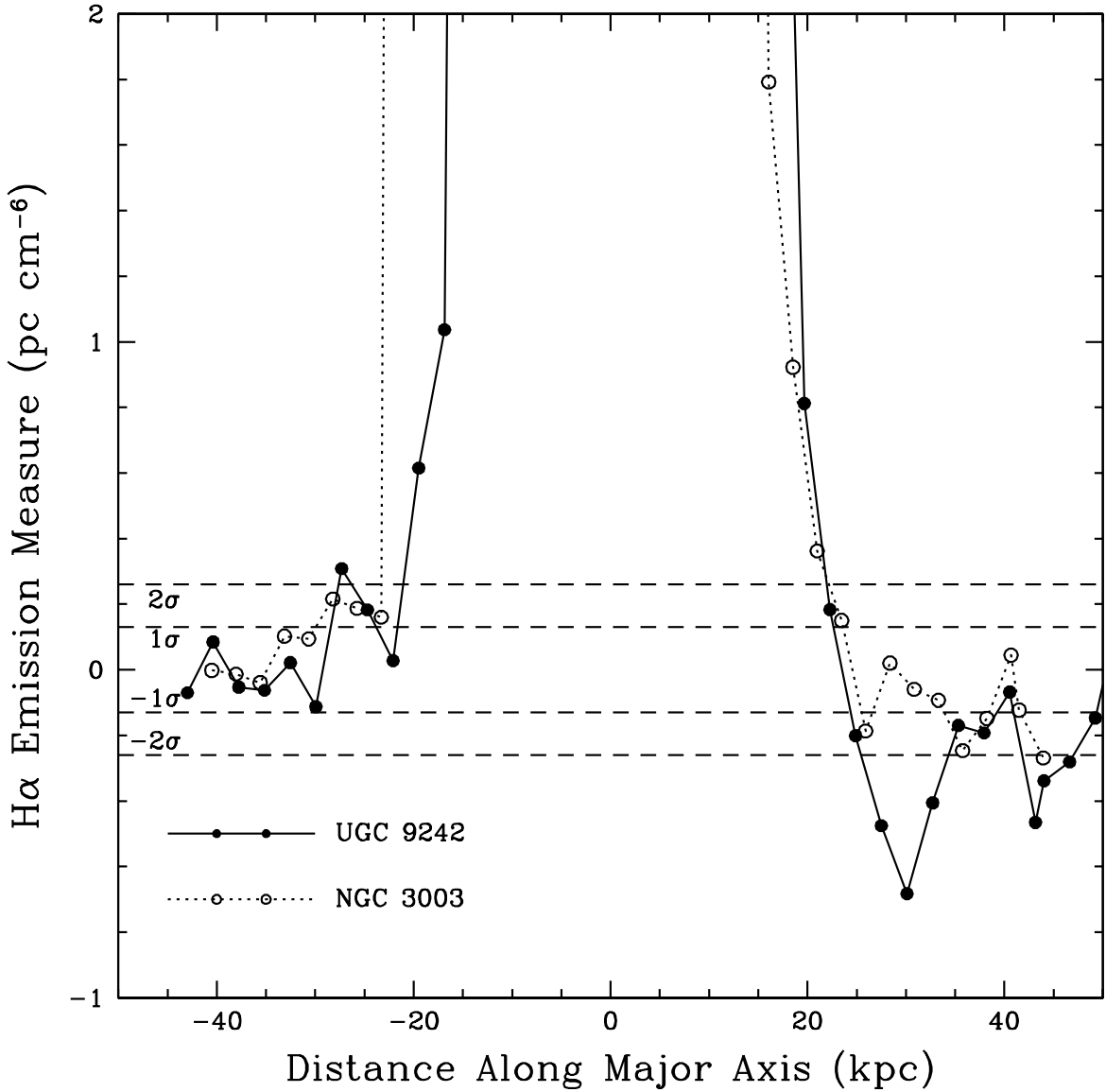}
\caption{The major axis profiles of UGC 9242 (solid line) and 
NGC 3003 (dotted line). The images were binned in 2.6 kpc $\times$ 2.6
kpc boxes for UGC 9242 and 2 kpc $\times$ 2 kpc boxes for NGC 3003, in
accordance with the limits on flatfielding accuracy found in section
6. The 1 and 2$\sigma$ flatfielding accuracy levels are also shown.}
\end{figure}

\newpage
\begin{figure}
\figurenum{15}
\epsscale{0.9}
\caption{A median filtered image of UGC 9242
(bottom panel) compared with the original H$\alpha$+continuum image
(top panel). The median box used was 20.7\arcs$\times$20.7\arcs. Halo
emission on both sides of the disk is apparent. Scattered light from a
bright star below the disk may be a problem, but no such problem
exists on the north side. The upper image shows the galaxy responsible
for the dark smudge in the upper left corner of the image, as well as
other galaxies and foreground stars that show up in the smoothed
image. Features in this image are discussed in more detail in the
text.}
\end{figure} 

\end{document}